# Unveiling the link between logical fallacies and web persuasion

**Antonio Lieto**
University of Turin, Dept. of Computer Science
Corso Svizzera 185
10149, Torino, Italy
lieto@di.unito.it

**Fabiana Vernero**
University of Turin, Dept. of Computer Science
Corso Svizzera 185
10149, Torino, Italy
vernerof@di.unito.it

## Abstract
In the last decade Human-Computer Interaction (HCI) has started to focus attention on persuasive technologies having the goal of changing users' behavior and attitudes according to a predefined direction. In this work, we hypothesize a strong connection between logical fallacies (forms of reasoning which are logically invalid but cognitively effective) and some common persuasion strategies adopted within web technologies. With the aim of empirically evaluating our hypothesis, we carried out a pilot study on a sample of 150 e-commerce websites.

## ACM Classification Keywords
H.5.m. Information interfaces and presentation (e.g., HCI): Miscellaneous.

## Introduction
In the last decade several studies in the field of Human-Computer Interaction (HCI) have started to focus attention on forms of persuasive interaction where the goal of one of the two agents involved in the process, namely, the artifact, is that one of "orienting" the attitudes and/or behaviors of the other agent (the user) according to a predefined direction. In this work, we hypothesize that there is a strong connection between logical fallacies (forms of reasoning which are logically invalid but cognitively effective, studied since the antiquity) and some of the most common persuasion strategies adopted within web technologies. With the aim of empirically evaluating our hypothesis, we present the results of a pilot study carried out on a sample of 150 e-commerce websites. The rest of the paper is organized as follows: Section 2 introduces the theme of fallacies; Section 3 presents a short introduction to captology (the research area which studies persuasive technologies); Section 4 is dedicated to the description of the connections we identified between fallacious arguments and some of the techniques used in persuasive technologies; then, Section 5 presents the methodology adopted for our pilot study and shows the obtained results. Some preliminary conclusions (Section 6) follow.

## Valid and Invalid Arguments

Logic[1] is "the discipline studying the theory of valid inferences" [10]. An inference[2] is composed by a set of initial propositions (*premises*) from which other propositions (*conclusions*) are derived. All the valid rules of classical logic are based on *deductive* inferential schemes where the conclusion C is a logical consequence of a set of premises P1...Pn. An example of deductive inference is the following:

P1: All the men are mortal
P2: Socrates is a man
C: Socrates is mortal

---

[1] Here with this term we refer to classical formal logic.

[2] For the sake of simplicity, we will consider here the term "inference" as a synonym of the term "argument".

However, not all the inferences are deductive and, therefore, logically valid [4]. There are, in fact, several types of *inductive*[3] inferences where the conclusion does not logically follow from the premises. An example of inductive inference is shown below:

P: All the Italians that I know are lazy
C: Italians are lazy

Within the class of inductive inferences, logical fallacies enjoy a special status. In fact, they are inferences that, "even if invalid from a formal point of view[4], appear as plausible and therefore are psychologically persuasive" [4][9]. According to this definition, not all inductive inferences can be considered as fallacious.

An important aspect to point out regards the connection between inferential validity and rationality. From our point of view, a fallacious argument does not necessarily is "irrational". Indeed, since the psychological/cognitive aspect plays a crucial role in the dynamics of persuasion, a fallacious argument is usually an invalid argument endowed of psychological plausibility and a proper heuristic value.

---

[3] For the sake of simplicity, here we will refer to all the inferences that are not deductive with the term "inductive inference". Therefore even the abduction, in this case, can be ascribed to the category of "inductive inferences".

[4] In the field of argumentation theory a number of criticisms have been raised about the use of classical logic as an instrument for the analysis of fallacious arguments, and some alternative solutions have been proposed in order to justify the use of such arguments in certain contexts (e.g. in the case of the "New Dialectic" approach proposed by Douglas Walton [13]). However such criticisms have, in our opinion, some limits, as already pointed out in [3]. More specifically: i) they do not allow to characterize the difference between *fallacies*, *errors*, and *weak arguments*, (ii) the risk of "relativism" seems to be around the corner since these approaches hypothesize contexts where the traditional fallacies are no more considered "fallacious".

The study and classification of logical fallacies goes back to the Philosopher in the *De Sophistichis Elenchis* [2]. During the centuries different research areas such as logic, rhetoric and argumentation theory dealt with this problem, pointing out that fallacious arguments are suitable to be used as techniques for achieving persuasive goals [12].

## Captology

In the Nineteen Nineties, B.J. Fogg [8] coined the term "captology", as an acronym for the expression "computers as persuasive technologies", to describe a research area which regards computer technologies as potential persuaders and focuses on both their analysis and their design. According to Fogg, persuasion can be defined as an attempt "to change attitudes or behaviors or both" [8]. Following on from this definition, all computer technologies which are purposely designed with the aim of changing their users' attitudes or behaviors can be considered as persuasive [8].

## Fallacies and Persuasive Technologies

In the field of captology, the above mentioned connection between fallacies and persuasion has never been - at the best of our knowledge – investigated and pointed out. In our opinion, however, most of the techniques used in persuasive tech environments are, unknowingly, based on fallacious arguments. In the rest of this section we will present the connections that we identified between some well-known logical fallacies and some of the techniques used in the field of persuasive technologies.

The logical fallacy known as *argumentum ad populum*, or "appeal to the majority", consists in accepting a certain thesis based on the mere fact that most people accept it. A typical example of such a fallacy is: "Most people like a certain book, then that book is worth-reading". This fallacy can be compared to those strategies, commonly used in the realm of persuasive technologies, which owe their persuasive potential to the exploitation of social dynamics. For example, technologies which grant access to social networks can leverage influence dynamics among peers to stimulate their users to attain certain goals. More specifically, Fogg refers to well-known social psychology theories, such as *social comparison* and *conformity* [13], which can be applied to computer technologies. According to social comparison, people who are uncertain about the way they should behave in a situation actively seek information about others and use such information to form their own attitudes and behaviors. Conformity theory, on the contrary, focuses on normative influence, claiming that people who are part of a group usually experience a pressure to conform to the expectations of the other members of their group.

A further commonality with fallacies can be found focusing on the discussion about *credibility* which characterizes the area of persuasive technologies [7]. The perceived credibility (and, therefore, persuasiveness) of both people and computers is known to be affected by the so-called *halo effect* [5], according to which a positive evaluation with respect to a certain feature (e.g., physical attractiveness) produces a "halo" which causes an extension of such an evaluation to other, unrelated, features (e.g., expertise). Similarly, the fallacy of *argumentum ad verecundiam* (also "appeal to authority") refers to cases where some theses are assumed to hold based on the fact that the person asserting them is wrongly assumed to be an authority about the topic of the discourse because of his/her achievements in other, unrelated, fields. An example of such a fallacious argument is the following: "the economist X claims that vegan diet is dangerous for our health. Therefore: it is wrong to follow vegan diets".

Technologies which implement tailoring techniques are persuasive because they provide each individual with the information they are likely to find the most interesting, based on their personal preferences, goals and experience. Obtaining personalized information does not only save users the effort to examine an overwhelming amount of content, but it is also more likely to draw their attention and, in case the so-obtained information is accepted, it can determine deeper and longer-lasting changes. Various personalization techniques are commonly adopted in adaptive systems and in recommender systems, such as collaborative and content-based filtering [1]. Personalization techniques can be considered fallacious because they are based on the assumption that (i) people will maintain their past preferences in the future (content-based filtering) or that (ii) people who have proved to have similar preferences in the past will maintain this similarity also in the future (collaborative filtering), which, although being probable, cannot be taken for granted. Tailoring can be compared to the so-called "*audience agreement*" technique, which is well known in rhetoric and theory of argumentation [12]. According to this technique, persuaders should only use arguments which have already been accepted by their audience.

Furthermore, surveillance technique can be compared to the *argumentum ad baculum* fallacy. Surveillance is based on the idea that people tend to change the way they behave when they are aware that they are being observed, especially if the observer has the power to punish or reward them (in this case, they will tend to match the observer's expectations) [13]. The covert menace which underlies surveillance technique is not too dissimilar to the argumentum ad baculum, where the persuader resorts to threats of force in order to make his/her thesis be accepted. An example of this fallacy, inspired to Pascal's Gamble [11], is: "If you don't believe that God exists, when you die you will be judged and sent to Hell, so it is safer to believe in God". The argumentum ad baculum plays a central role in another fallacy, known as *argumentum ad consequentiam*, according to which a proposition is accepted based on the desirability or undesirability of its consequences (a positive example of this fallacy is: "If there is an afterlife, then we will meet our loved ones again. Therefore: there must be an afterlife"). In the field of persuasive technologies, allowing users to explore cause-and-effect relationships is a well-known technique, which exploits the possibility to offer computer simulations where users can manipulate certain inputs (e.g., their daily food intake) and observe their consequences (e.g., their weight) [8]. Prominent examples which show how cause-and-effect simulations can be used with persuasive effects can be found in environmentalist websites which allow users to calculate their ecological footprint (i.e., the number of planets which would be needed if everyone lived like them) based on their lifestyle and consumption habits.

| Fallacy | Website features |
|---|---|
| Arg. ad populum | Best sellers, ratings |
| Arg. ad verecundiam | Improper testimonials |
| Audience agreement | Personalization |
| Arg. ad baculum | Visibility of purchased browsed items or wish lists |
| Arg. ad consequentiam | Cause-effect simulations |
| Accent | Emphasis/hiding of information |

**Table 1.** Correspondence table fallacies/website features.

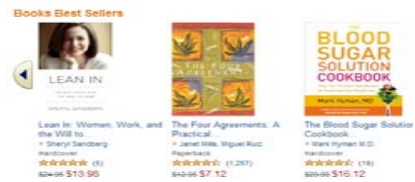

**Fig. 1**. Example of Argumentum ad Populum in the Amazon website.

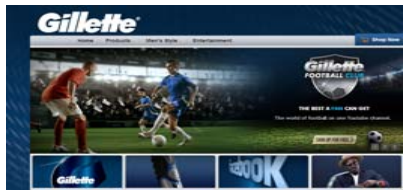

**Fig. 2**. Ex. Argumentum ad Verecundiam in the Gillette website (e.g. the testimonials are the football players).

Finally, the *accent* fallacy, which occurs when emphasis is used to manipulate the actual meaning of a proposition, is commonly adopted with a persuasive intent in computer technologies, especially in its visual variant where certain elements are made more visually prominent in order to emphasize them. A common example of the (visual) accent fallacy occurs when special offers (e.g., discounts) are highlighted with big fonts and bright colors, while the possibly restrictive conditions to enjoy them are made scarcely visible. In Human-Computer Interaction, the accent fallacy can be compared to misplaced salience, which is known as one of the "demons" hindering situation awareness [6]. While appropriate salience can help to identify the most important information in a certain context, misplaced salience emphasizes irrelevant cues, confusing users and leading them to inappropriate behaviors.

## Pilot Study: Methodology and Evaluation

In order to evaluate our hypothesis, we carried out a pilot study on the first 150 e-commerce websites indexed by Google for the query "on-line shopping". Our basic assumption is that websites with a clear persuasive goal, such as e-commerce websites, adopt many persuasive techniques which are studied in the field of captology. With our evaluation, we aimed at investigating whether, among the techniques used in the persuasive technologies, there actually are some which are reducible to arguments based on logical fallacies, as shown in Section 4. In order to run the evaluation we created a correspondence table where the presence of fallacious arguments is connected to the use of some specific features in the websites.

As shown in Table 1, the argumentum ad populum has been associated to the case in which either "best seller" products (fig. 1) or user ratings are displayed (in this case the persuasion strategy is based on the following argument: "Most people buy/like X, then it is positive to buy X"); the argumentum ad verecundiam has been associated to the presence, in one of more parts of the website, of improper testimonials for certain products (see e.g. fig. 2); the "audience agreement" has been associated with the use of recommendation techniques, and the argumentum ad baculum with the presence of software environments that make the actions performed on the website totally "transparent", so that users are induced to buy products (or services) which are consistent with the self-image they want to show to others. Finally, the argumentum ad consequentiam has been associated to the presence of software environments able to simulate the consequences of certain user choices[5], and the accent fallacy to the case when part of the purchasing-related information is emphasized and part is hidden (e.g. when shipping or tax costs are presented only at the end of the purchasing process). Table 2 shows the obtained results.

| Fallacy | % |
|---|---|
| Arg. ad populum | 50 % |
| Arg. ad verecundiam | 15,3 % |
| Audience agreement | 32 % |
| Arg. ad baculum | 4 % |
| Arg. ad consequentiam | 9,3 % |
| Accent | 56 % |

---

[5] Since the decision about which premises (e.g. the input of the users) drive to a "good" consequence and which to a "bad" one is predetermined by the website designers, this technique has a clear persuasive goal.

**Table 2.** Percentage of websites using fallacious-reducible persuasive mechanisms. The complete list of data derived from the websites is at: https://sites.google.com/site/techsuasion/

## Preliminary Conclusions and Future Work

The results of the pilot study show that arguments based on logical fallacies are widely used within e-commerce websites. However, since in the literature about persuasive technologies there is no reference about the connection between fallacies and persuasive techniques, we believe that the use of fallacious-reducible persuasive mechanisms is essentially done unknowingly. The main contribution of this work is, therefore, that one of unveiling this link (at least for web-based persuasive environments such as e-commerce websites). This unveilment, in our opinion, can provide the research about web persuasion and captology with a huge theoretical basin (namely that one provided by the disciplines that, during the centuries, dealt with all the major aspects of logical fallacies) that could be a possible source of inspiration for the study and design of "computer-driven" persuasion mechanisms. In our opinion, in fact, acknowledging and exploiting this theoretical root could improve the efficacy of persuasive technologies. In our future work we plan to extend our analysis by increasing the number of both the logical fallacies examined and the technological environments where they have been (or can be) used. Furthermore we aim at extending the correspondence matrix between fallacies and persuasion techniques. Beyond the goal of creating a "*catalogue raisonné*", this study will also allow to point out which logical fallacies, not yet finding a correspondence in the field of captology, could be used as a basis for the design of novel persuasive technologies.